\documentclass{iaus}
\usepackage{graphicx}
\usepackage{subfigure}
\usepackage{natbib}

\def\aap{\mbox{\bf{Astron.\ Astrophys.}}}
\def\apj{\mbox{\bf{Astrophys.\ J.}}}

\def\pasp{\mbox{\bf{Pub.\ Astron.\ Soc.\ Pacific}}}

\def\sci{\mbox{\bf{Science}}}
\def\araa{\mbox{\bf{Ann.\ Rev.\ Astron.\ Astrophys}}}

\def\grl{\mbox{\bf{Geophysical\ Research\ Letters}}}

 \def\mso{\,\mathrm{M}_\odot}

 \def\kms{\, {\rm km}\, {\rm s}^{-1}}

 \def\simle{\mathrel{\hbox{\rlap{\hbox{\lower4pt\hbox{$\sim$}}}\hbox{$<$}}}}
 \def\simgr{\mathrel{\hbox{\rlap{\hbox{\lower4pt\hbox{$\sim$}}}\hbox{$>$}}}}

 \def\grad{\nabla}
 \def\adgrad{\nabla_{\mathrm{\!ad}}}
 \def\mugrad{\nabla_{\!\mu}}
 \def\ath{\alpha_{\mathrm{th}}}

 \def\c2{^{12}{\mathrm C}}
 \def\c3{^{13}{\mathrm C}}
 \def\n14{^{14}{\mathrm N}}
 \def\c1213{^{12}{\mathrm C}/^{13}{\mathrm C}}
 \def\he3he4{^3\mathrm{He}/^4\mathrm{He}}

\title[Thermohaline mixing in low-mass giants]{Thermohaline mixing in low-mass giants}

\author[M. Cantiello \& N. Langer]
{M. Cantiello$^1$ \and  N. Langer$^1$}

\affiliation{$^1$Astronomical Institute Utrecht,
              Princetonplein 5, 3584 CC Utrecht, The Netherlands\break           
	      email: M.Cantiello@uu.nl, N.Langer@uu.nl
}

\pubyear{2008}
\volume{252}  
\pagerange{???-???}
\date{?? and in revised form ??}
\jname{Proceedings The Art of Modelling Stars in the 21st century}

\editors{Licai Deng  K.L. Chan \& C. Chiosi, eds.}

\begin{document}

\maketitle

\begin{abstract}
Thermohaline mixing has recently been proposed to occur in
low mass red giants, with large consequences for the chemical yields
of low mass stars. We investigate the role of thermohaline mixing during the evolution
of stars between 1$\mso$ and 3$\mso$, in comparison to other mixing processes
acting in these stars. We confirm that thermohaline mixing has the potential to destroy most of
the $^3$He which is produced earlier on the main sequence during the red giant
stage. In our models we find that this process is working only in stars with initial mass  $M \simle1.5\mso$. 
Moreover, we report that thermohaline mixing can be present during core helium
burning and beyond in stars which still have a $^3$He reservoir. While rotational and magnetic mixing is negligible compared to
the thermohaline mixing in the relevant layers, the interaction of thermohaline 
motions with differential rotation and magnetic fields may be essential to establish the time scale
of thermohaline mixing in red giants. 

\keywords{ stars: evolution, stars: abundances, stars: AGB and post-AGB, stars: magnetic fields, stars: rotation, ISM: abundances }

\end{abstract}
\firstsection 


\section{Introduction}



Thermohaline mixing is not usually considered as an important mixing process
in single stars, since the ashes of thermonuclear fusion consists of heavier nuclei
than its fuel, and stars usually burn from the inside out. The condition for
thermohaline mixing, however, is that the mean molecular weight ($\mu$)
decreases inward. 
Recently \citet[CZ07]{cz07} identified thermohaline mixing 
as an important mixing process which significantly modifies the surface composition
of red giants after the first dredge-up. The work by CZ07 was triggered by the paper
of \citet[EDL06]{edl06}, who found a $\mu$-inversion in their $1\mso$ stellar evolution model, occurring
after the so-called luminosity bump on the red giant branch (RGB), which is produced
after the first dredge-up, when the hydrogen-burning shell reaches the
chemically homogeneous part of the envelope. The $\mu$-inversion is produced by the
reaction $^3$He($^3$He,2p)$^4$He (as predicted by \citet{ulr72}). It does not show up earlier,
since the magnitude of the $\mu$-inversion is small and negligible compared to a stabilizing
$\mu$-stratification.

The mixing process below the convective envelope in models of low-mass stars turns out
to be essential for the prediction of the chemical yield of $^3$He (EDL06), and
to understand the surface abundances of red giants, in particular the $^{12}$C/$^{13}$C ratio, 
and the $^7$Li, carbon and nitrogen
abundances (CZ07).
This may also be important for other occurrences of thermohaline mixing in stars,
i.e., in single stars when a $\mu$-inversion is produced by off-center ignition
in semi-degenerate cores, or in stars which accrete chemically enriched matter 
from a companion in a close binary \citep[e.g.,][]{sgi+07}. Accreted  metal-rich matter
during the phases of planetary formation also leads to thermohaline mixing. 
The host stars of exoplanets present a metallicity excess compared to stars in which no planets have been detected.
This metallicity excess can be reconciled with the overabundances expected in cases of accretion
if thermohaline mixing is included in the picture \citep{vau04}.

\section{Method}
We use a 1-D hydrodynamic stellar evolution code \citep[][and references therein]{Yln06}. 
Mixing is treated as a diffusive process , the contributions to the diffusion coefficient are convection, semiconvection, 
thermohaline mixing, rotationally induced mixing and magnetic diffusion.
The code includes the effect of centrifugal force on the stellar structure, and
the transport of angular momentum is also treated as a diffusive process \citep{es78,Pks+89}.
The condition for the occurrence of thermohaline mixing is
\begin{equation}
\frac{\varphi}{\delta} \,\mugrad \le \grad - \adgrad \le 0 \label{condition}
\end{equation}
i.e. the instability operates in regions that are stable against convection (according to the 
Ledoux criterion) and where an inversion in the mean molecular weight is present. Here 
$\varphi =(\partial \ln \rho / \partial \ln
\mu )_{P,T}$, $\delta =-(\partial \ln \rho / \partial \ln T )_{P,\mu
  }$, $\nabla_\mu = d\ln\mu / d\ln P$, $\adgrad= (\partial \ln T/
\partial \ln P )_{\mathrm{\!ad}}$, and $\nabla = d\ln T / d\ln P$.
Numerically, we treat thermohaline mixing through a diffusion scheme
\citep{bra97,wlb01}. The corresponding diffusion coefficient is based on the
work of Stern (1960), \citet{ulr72}, and \citet{krt80}:
\begin{equation}
  D_{th} = -\ath\; \frac{3K}{2\,\rho\, c_P}\,
  \frac{\frac{\varphi}{\delta}\nabla_\mu}{(\adgrad - \grad)} \label{coefficient} 
\end{equation}
where $\rho $ is the density, $K=4acT^{3}/(3\kappa\rho )$ the thermal conductivity, and 
$ c_P=(dq/dT)_P$ the specific heat capacity. The quantity
$\ath$ is an efficiency parameter for the thermohaline mixing.
The value of this parameter 
depends on the geometry of the fingers arising from the instability and is still a matter of debate 
\citep{ulr72,krt80,cz07}. Most of the calculations have been performed
with $\ath=2.0$, corresponding to the prescription of \citet{krt80}, although we also investigated the effect of using different values of $\ath$.

In the code rotational mixing is included. Four different diffusion coefficients are calculated for dynamical shear, 
secular shear, Eddington-Sweet circulation and Goldreich-Schubert-Fricke instability. 
Details on the physics of these instabilities and their implementation in the code can be found in \citet{hlw00}.

Chemical mixing and transport of angular momentum due to magnetic fields \citep{spr02} is included as in \citet{hlw00}. 
The contribution of magnetic fields to the mixing is also calculated and added to the total diffusion
coefficient $D$ entering the diffusion equation.

We compute evolutionary models of $1.0\mso$, $1.5\mso$, $2.0\mso$ and $3.0\mso$ at solar metallicity (Z=0.02).
The initial equatorial velocities of these models were chosen to be 10, 45, 140 and 250 $\kms$ \citep{Tas00}; we assume
the stars are rigidly rotating at the zero-age main sequence. Throughout the evolution of all models, the mass-loss rate of 
\citet{rei75} was used.  

 \begin{figure}
 \begin{minipage}{6.85cm}
  \resizebox{\hsize}{!}{\includegraphics{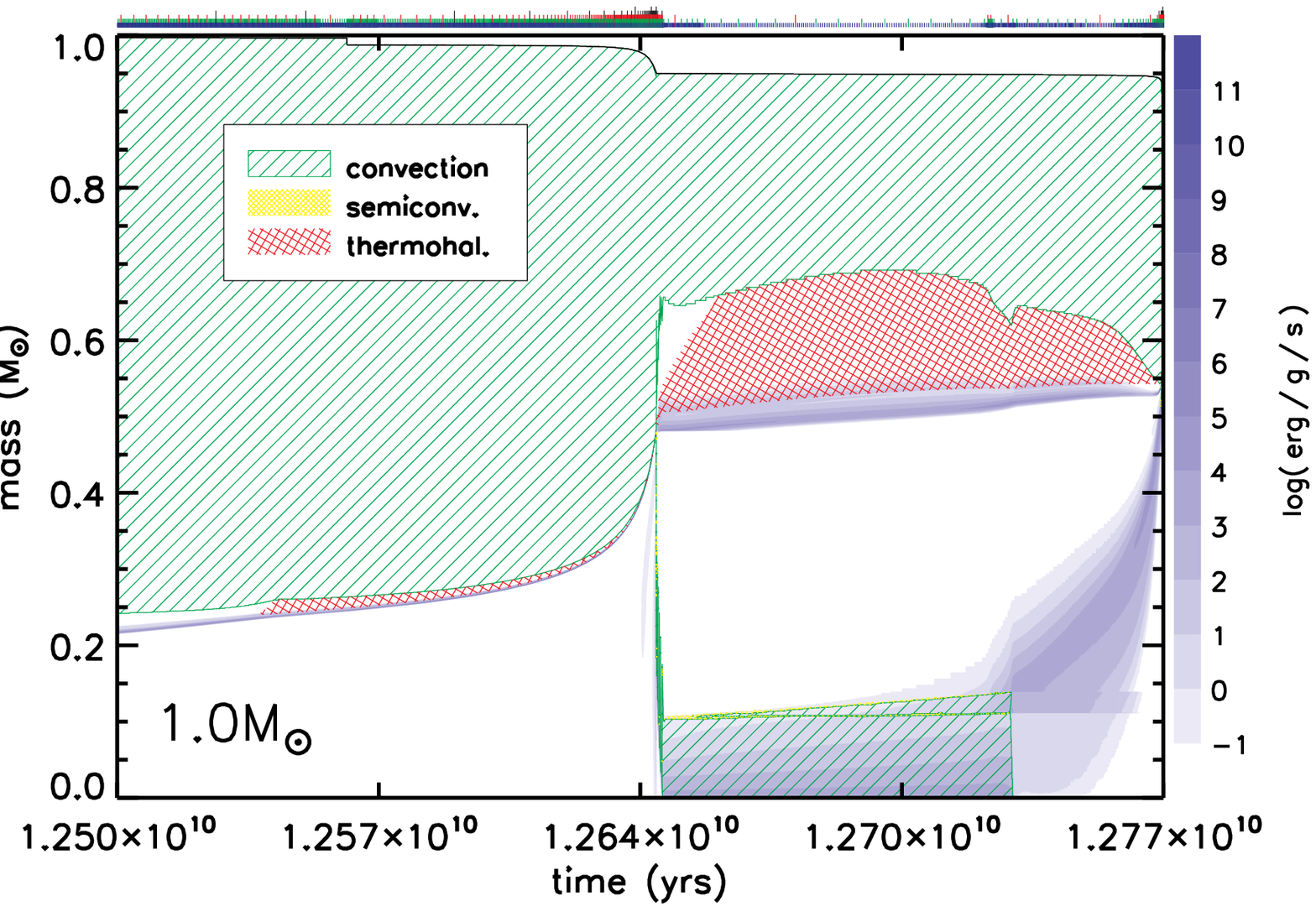}}
 \end{minipage}
 \hfill
 \begin{minipage}{6.85cm} 
   \resizebox{\hsize}{!}{\includegraphics{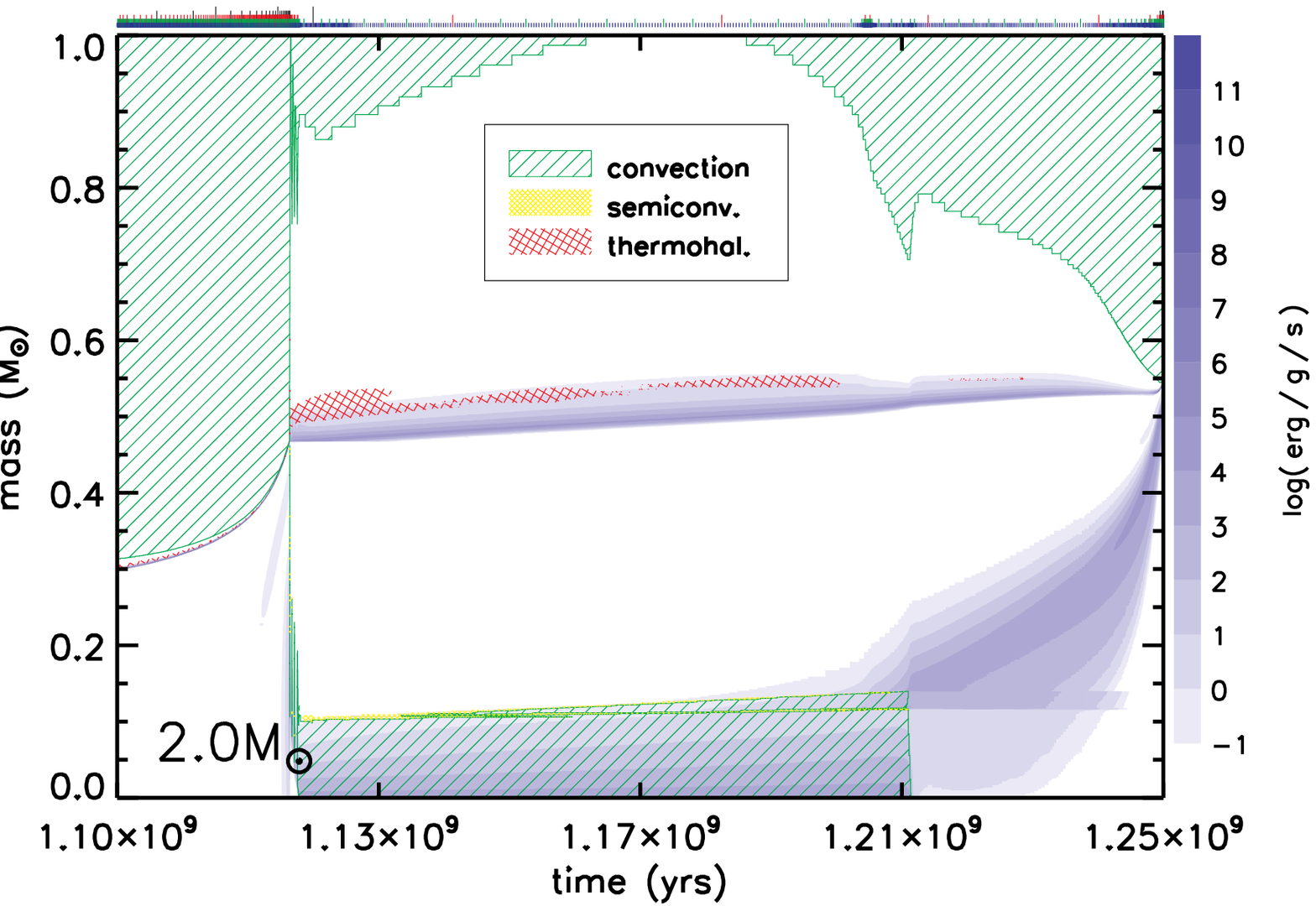}}
 \end{minipage}
\caption{{\bf Left: } evolution of the internal structure of a $1.0\mso$ star from the onset of thermohaline 
mixing to the AGB phase. Green hatched regions indicate convection, yellow
 filled regions represent semiconvection and red cross hatched regions indicate thermohaline mixing, 
as displayed in the legend. Blue shading shows regions of nuclear energy generation. {\bf Right: } same as left panel, but for a $2.0\mso$ star.}
\label{1e2}
 \end{figure}

\section{Results}

\begin{figure}
 \begin{minipage}{6.85cm}
  \resizebox{\hsize}{!}{\includegraphics{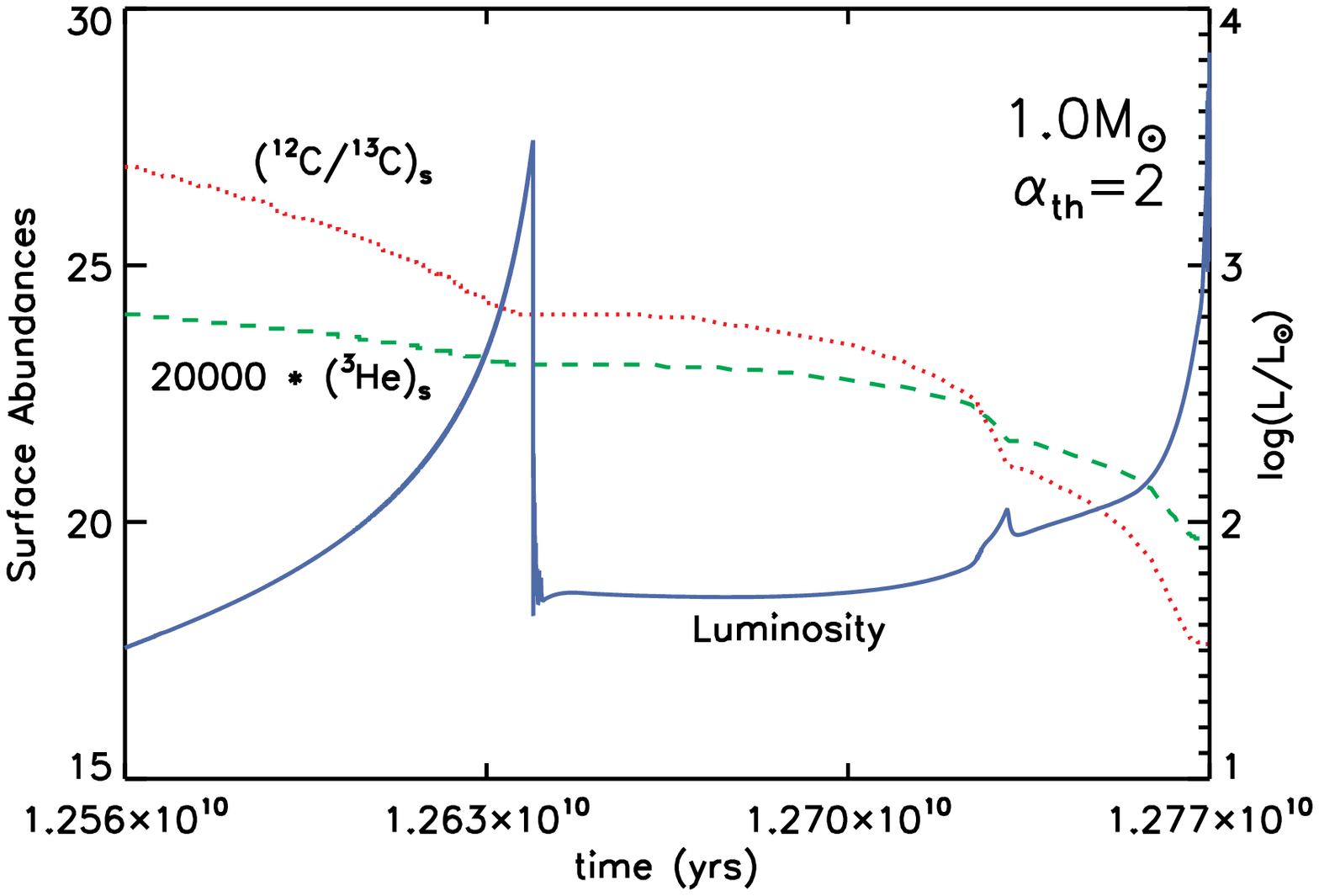}}
 \end{minipage}
 \hfill
 \begin{minipage}{6.85cm} 
   \resizebox{\hsize}{!}{\includegraphics{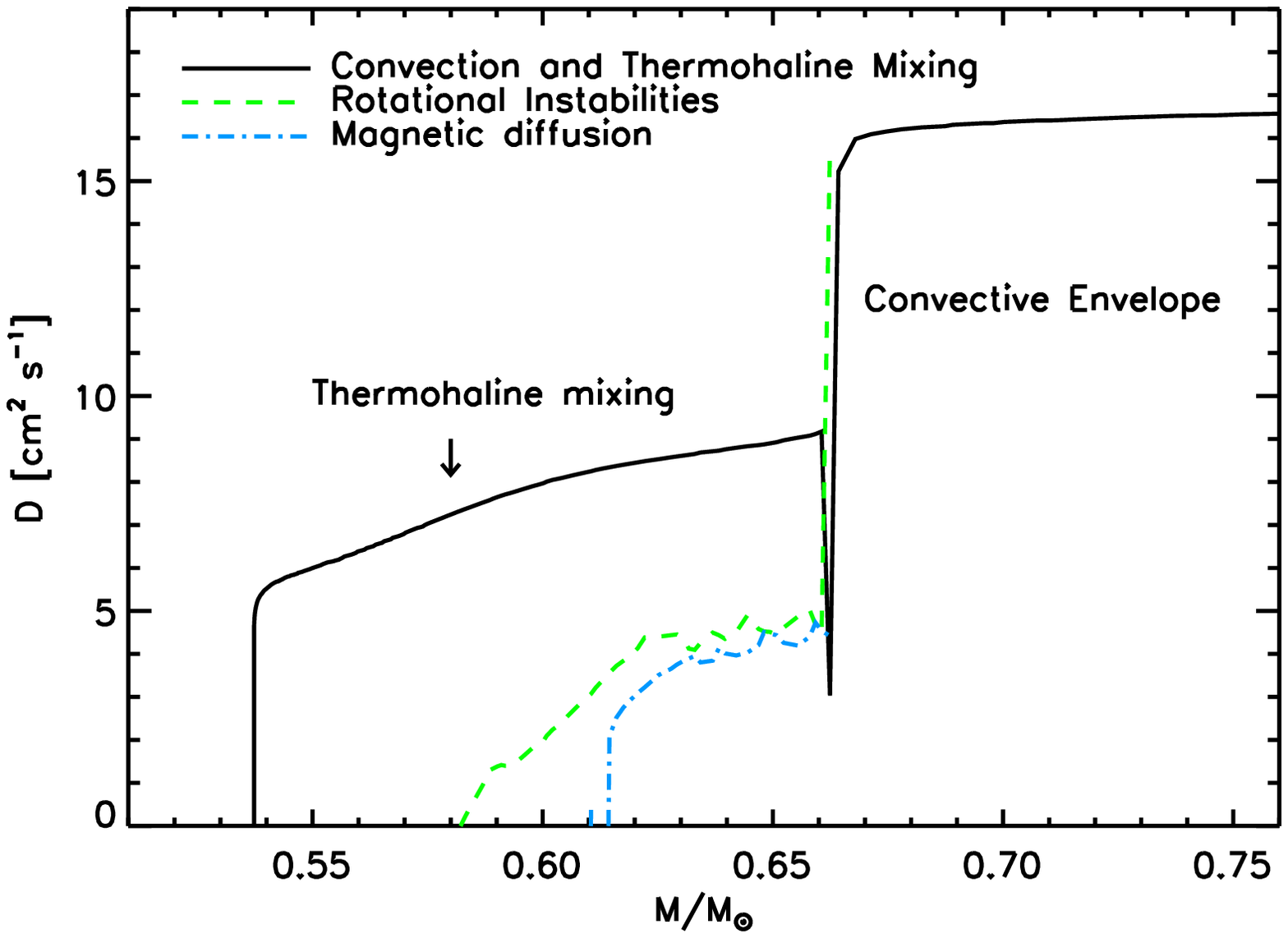}}
 \end{minipage}
\caption{{\bf Left: } evolution of the surface abundance of the $^{12}$C/$^{13}$C ratio 
(dotted red line) and $^3$He (dashed green line), and of the luminosity (solid blue line)
 from the onset of thermohaline mixing up to the AGB phase for a $1.0\mso$ star. {\bf Right: } diffusion coefficients 
in the region between the H burning shell and the convective envelope 
for the $1.0\mso$ model during core He-burning (t $=1.267\times10^{10}$ years). 
The black, continuous line shows convective and thermohaline mixing diffusion coefficients, the green, 
dashed line is the sum of the diffusion coefficients due to rotational 
instabilities while the blue, dot-dashed line shows the magnitude of magnetic diffusion coefficient.}
\label{surfacediff}
 \end{figure}

We confirm the presence of an inversion in the mean molecular weight in
the outer wing of the H-burning shell, after the luminosity bump on the red giant branch.
According to inequality (\ref{condition}) this inversion gives rise to  thermohaline mixing  in the radiative 
buffer layer, the radiative region between the H-burning shell and the convective envelope.\\
Results of our calculations for the 1.0 and 2.0 $\mso$ models are shown in Fig. \ref{1e2}.
In our 1$\mso$ model, thermohaline mixing develops at the luminosity bump and transports chemical species in  
the radiative layer between the H-burning shell and the convective envelope. This results 
in a change to the stellar surface abundances. The left panel of Fig. \ref{surfacediff} shows the evolution of the $^3$He surface abundance, 
and of the ratio $\c1213$ at the surface as a function of time in the 1$\mso$ model, confirming the result of EDL06 and CZ07, namely that thermohaline mixing
is efficiently depleting $^3$He and lowering the ratio $\c1213$ on the giant branch. 
Interestingly the 2$\mso$ model shows a different behaviour: thermohaline mixing is not able to connect the H-burning shell 
and the convective envelope, resulting in no change to the surface abundance of $^3$He and to the ratio $\c1213$ after the luminosity bump.
Our models with $\ath=2.0$ show thermohaline mixing to be important during the RGB only in stars with mass $M \simle1.5\mso$.

While CZ07 and EDL07 investigate thermohaline mixing only during the RGB phase, we followed the evolution of 
our models until the thermally pulsating AGB phase (TP-AGB).
In fact a $\mu$-inversion is always created if a H-burning shell is active in a chemically homogeneous layer; this  happens not only during the RGB phase, 
but also during the horizontal branch (HB) and the AGB. The size of the $\mu$-inversion depends on the local amount of $^3$He.
As a consequence, the efficiency of thermohaline mixing in different evolutionary phases depends on the amount of $^3$He left by previous 
mixing episodes.   

After the core He-flash, helium is burned in the core, while a H-burning shell is active below the convective envelope.
We found that during this phase thermohaline mixing is present and can spread through the whole radiative buffer layer in our 1$\mso$ model (left panel in Fig. \ref{1e2}). 
In this model the surface abundances change also during this phase because the H-burning shell and the envelope are connected. This is shown in  Fig. 
\ref{surfacediff}, left panel, where surface abundances change after the luminosity peak corresponding to the He-flash.
 We stress that using the prescription of \citet{krt80} for thermohaline mixing allows our model to reach this phase without completely burning the $^3$He; 
models of CZ07 almost completely 
deplete $^3$He in the envelope during 
the RGB phase because of their higher diffusion coefficient \citep{ulr72}.
In this case thermohaline mixing would be much less efficient, during the subsequent evolutionary phases, due to the lower abundance of $^3$He.

\begin{figure}
 \begin{minipage}{6.85cm}
  \resizebox{\hsize}{!}{\includegraphics{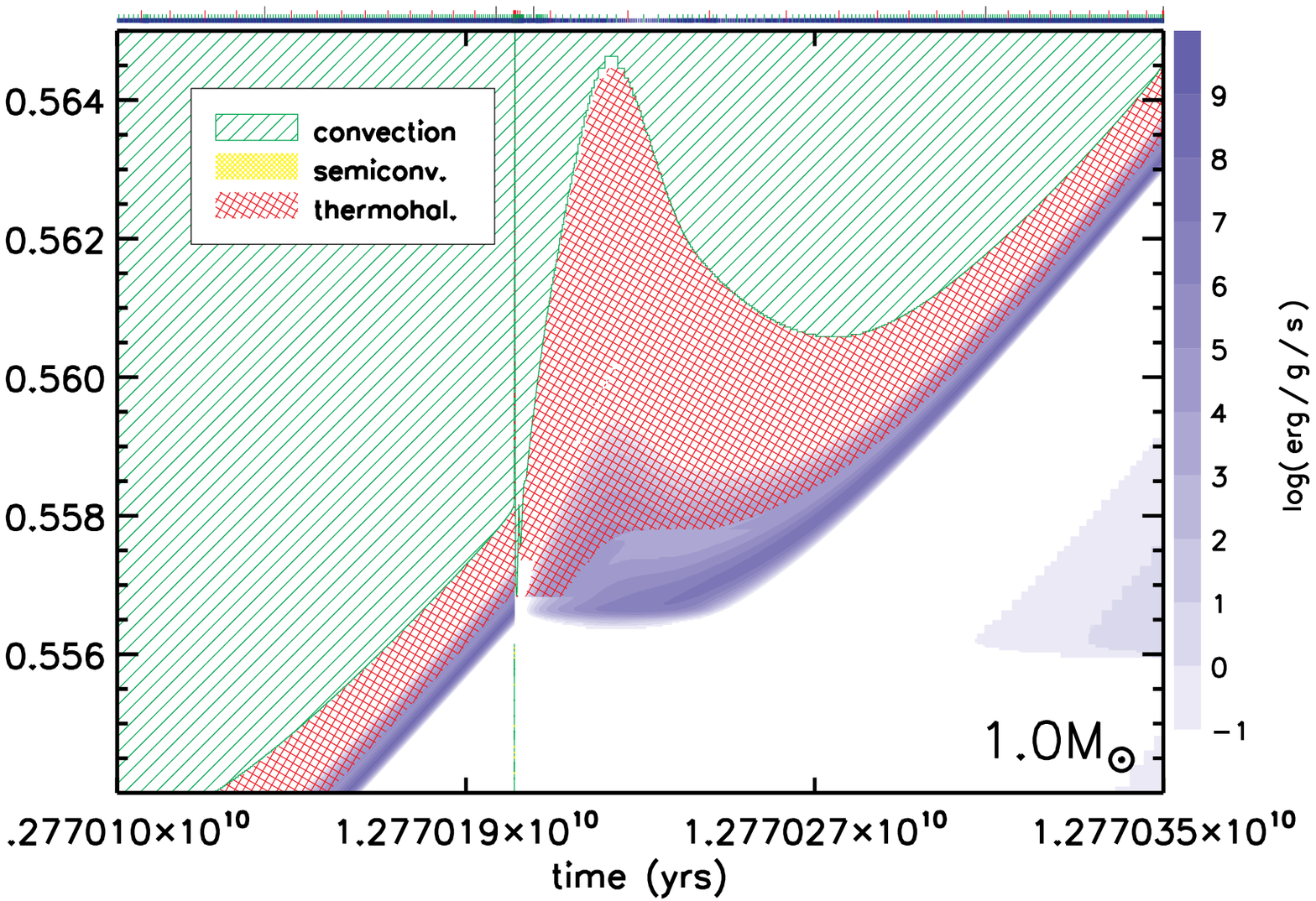}}
 \end{minipage}
 \hfill
 \begin{minipage}{6.85cm} 
   \resizebox{\hsize}{!}{\includegraphics{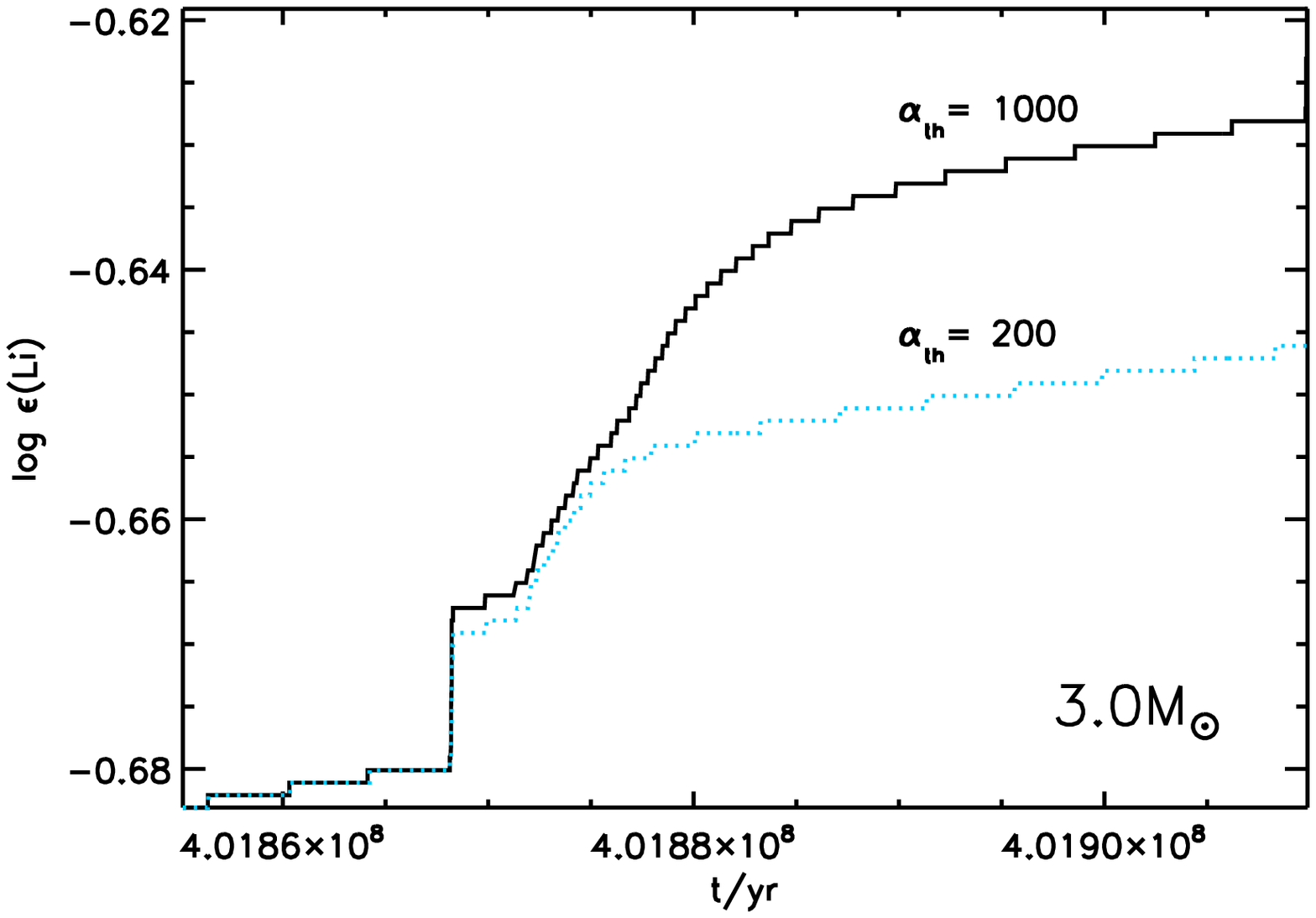}}
 \end{minipage}
 \caption{{\bf Left: }evolution of 
the region between the H burning shell source and the convective 
envelope during a thermal pulse in a $1.0\mso$ star. Green hatched regions indicate convection and red 
crossed regions indicate thermohaline mixing. Blue shading shows regions of nuclear energy generation. 
{\bf Right: }evolution of Lithium surface abundance during one thermal pulse in a 3$\mso$ model. The black, 
continuous line shows a model evolved with $\ath$=1000; the blue, dotted line refers to the same model 
evolved with a thermohaline mixing efficiency $\ath = $200.
 In both cases the model is experiencing third dredge-up. The evolution of the star prior to the TP-AGB has been 
calculated with $\ath =$ 2.}
\label{pulse}
\end{figure}

The last nuclear-burning phase of a low-mass star is the AGB, and is characterized by the presence of two burning shells and a degenerate core.
The star burns H in a shell and the ashes of this process feed an underlying He shell. 
During the most luminous part of the AGB the He shell periodically experiences thermal pulses (TPs); in stars more massive
than about 1.2$\mso$ these TPs are associated with a deep penetration of the convective envelope, the so-called third dredge-up (3DUP) \citep{wk98}.
We find thermohaline mixing to be present during the TP-AGB phase. Depending on the mass of the model the diffusion process is able to connect the
 H-burning shell with the convective envelope during the whole interpulse phase.
In a $1\mso$ model thermohaline mixing connects the H-burning shell 
to the convective envelope (Fig. \ref{pulse}, left panel), confirming that this mixing process is more efficient at lower masses. 
The occurrence of thermohaline mixing in this late evolutionary phase is
critically determined by the mixing history of the star. This is because the amount of  $^3$He present during the TP-AGB phase is controlled by
the occurrence of thermohaline mixing and by its efficiency ($\ath$) in previous evolutionary phases.

Thermohaline mixing during the TP-AGB is a circulation of the type inferred by Cameron \& Fowler (1971), since it carries $^7$Be out of the high-temperature
zone. $^7$Be can then produce $^7$Li by electron capture, which is transported in the convective envelope. This results in surface enrichment of $^7$Li, 
as shown in the right panel of Fig. \ref{pulse}.
Interestingly \citet{ulp07} recently reported the detection of low-mass, Li-rich AGB stars in the galactic bulge. Given their low mass, these stars are not expected to 
experience any hot bottom burning. We argue that thermohaline mixing is a possible explanation for their abundance anomalies.

\section{Rotation and magnetic fields}
In our models we found that in the relevant layers thermohaline mixing
has generally higher diffusion coefficients than rotational instabilities and magnetic diffusion. This result is not valid for
magnetic stars, stars that possess anomalous surface fields of a few $10^2$ to about $10^4$ G, believed to be of fossil origin \citep{cz07b}.   
The right panel of Fig. \ref{surfacediff} clearly shows that rotational and magnetic mixing are negligible compared to
the thermohaline mixing in our $1.0\mso$ model. The only rotational instability acting on a shorter
timescale is the dynamical shear instability, visible in the right panel of Fig. \ref{surfacediff} (dashed line) as a spike  present at  the lower 
boundary of the convective envelope.
This instability works on the dynamical timescale in regions of a star characterized by a high degree of differential rotation. 
However, if present, this instability acts only in a 
very small region (in mass coordinate) at the bottom of the convective envelope. As a result thermohaline mixing is still setting the
timescale for the diffusion of chemical species from the convective envelope to the hydrogen-burning shell.

The interaction between rotation and thermohaline mixing is more difficult to address, since it requires full hydrodynamic calculations.
We expect the speed of thermohaline mixing to be affected by  differential rotation and rotational instabilities, since these are able to change 
the geometry of the fingers, and to create turbulence, respectively. Canuto et al. (2008) recently suggested that turbulence must be taken into account
to explain both laboratory and ocean data of double diffusive processes.    

The interaction of magnetic fields with thermohaline mixing must also be considered. This depends on the geometry and the magnitude of the magnetic field: 
if the magnetic field is stronger or of the same order of the equipartition value the magnetic
field allows plasma motions only along the direction of the field lines. This could result in the inhibition of thermohaline mixing, as discussed with detailed 
calculations by Zahn and Charbonnel (2008). However strong magnetic fields are observed in only a small fraction of low-mass stars (Wolff (1968); Power et al. (2007)). 
Concerning the interaction of weak fields with thermohaline motions, a change in the 
speed of the fingers is also expected and fully MHD calculations are needed to understand the process.


\section{Discussion}
We comfirm the results of EDL06 and CL07: thermohaline mixing in low-mass giants is capable
of destroying large quantities of ${^3}$He, as well as decreasing the ratio $\c1213$.
Thermohaline mixing starts when the hydrogen burning
shell moves into the chemically homogeneous layers established by the first dredge-up. 
Our models show further that thermohaline mixing remains important during core helium burning,
and can still be relevant during the AGB phase --- including the termally-pulsing AGB stage.   
This can result in important changes to the surface abundances of low-mass stars. The quantitative discussion is complicated
by the fact that the capability of thermohaline mixing to change surface abundances depends on an uncertain efficiency parameter, as well as on the local ${^3}$He abundance. 
The efficiency parameter $\ath$ is still a matter of debate and is probably strongly affected by the interaction of thermohaline mixing with rotation and magnetic fields.
The local ${^3}$He abundance depends on the previous history of mixing. 

Our calculations show that in the relevant layers, thermohaline mixing generally has a higher diffusion coefficient than rotational instabilities
and magnetic diffusion. Still, the interaction of thermohaline mixing with magneto-rotational instabilities is important: we expect the speed of the mixing to be strongly 
affected by the presence of differential rotation and magnetic fields. To better understand the picture it would be desirable to have realistic MHD simulations of thermohaline mixing. 


\begin{acknowledgments}MC wishes to thank V.Canuto, J.-P.Zahn and C.Charbonnel for useful discussions. MC acknowledges LKBF and the IAU for financial support. 
\end{acknowledgments}


\end{document}